\documentclass[journal]{IEEEtran}

\ifCLASSINFOpdf
\else
   \usepackage[dvips]{graphicx}
\fi
\usepackage{url}
\usepackage{amsfonts}
\usepackage{multirow}
\usepackage{graphicx}

\begin{document}

\title{Audio-visual Speech Separation with Adversarially Disentangled Visual Representation}

\author{Peng Zhang, Jiaming Xu, Jing Shi, Yunzhe Hao, and Bo Xu, \IEEEmembership{Member, IEEE}
\thanks{ }
\thanks{P. Zhang, J. Xu, J. Shi and Y. Hao are with the Institute of Automation, Chinese Academy of Sciences (CASIA), Beijing, China, and University of Chinese Academy of Sciences, China. (e-mail: zhangpeng2018@ia.ac.cn; jiaming.xu@ia.ac.cn; shijing2014@ia.ac.cn; haoyunzhe2017@ia.ac.cn).}
\thanks{B. Xu is with the Institute of Automation, Chinese Academy of Sciences (CASIA), Beijing, China, University of Chinese Academy of Sciences, China, and Center for Excellence in Brain Science and Intelligence Technology, CAS, China. (e-mail: xubo@ia.ac.cn).}}

\markboth{Journal of \LaTeX\ Class Files}
{Shell \MakeLowercase{\textit{et al.}}: Bare Demo of IEEEtran.cls for IEEE Journals}
\maketitle

\begin{abstract}
Speech separation aims to separate individual voice from an audio mixture of multiple simultaneous talkers. Although audio-only approaches achieve satisfactory performance, they build on a strategy to handle the predefined conditions, limiting their application in the complex auditory scene. Towards the cocktail party problem, we propose a novel audio-visual speech separation model. In our model, we use the face detector to detect the number of speakers in the scene and use visual information to avoid the permutation problem. To improve our model's generalization ability to unknown speakers, we extract speech-related visual features from visual inputs explicitly by the adversarially disentangled method, and use this feature to assist speech separation. Besides, the time-domain approach is adopted, which could avoid the phase reconstruction problem existing in the time-frequency domain models. To compare our model's performance with other models, we create two benchmark datasets of 2-speaker mixture from GRID and TCD-TIMIT audio-visual datasets. Through a series of experiments, our proposed model is shown to outperform the state-of-the-art audio-only model and three audio-visual models.
\end{abstract}


\IEEEpeerreviewmaketitle

\section{Introduction}
\IEEEPARstart{R}{ecent} deep learning approaches have made significant progress in the speech separation task, which is also famously known as the Cocktail Party Problem [1], [2]. However, there remain many unresolved issues, such as permutation problem [3] and an unknown number of sources in the mixture. Although many proposed audio-only single-channel methods attempt to alleviate these problems and achieve promising results [3]-[6], these works are built on a strategy to handle the predefined conditions, such as the number of speakers is determined.

In a cocktail party, human listeners pay attention to a target speaker by multiple cues from different modalities [7]. The visual cue is one of the most powerful and robust of these cues. Thus, it is a natural idea to leverage both the audio and visual cues to design a computational auditory model. Using visual cue, two problems mentioned in the last paragraph could be resolved in a unified model elegantly. First, we use a face detector to determine the number of speakers in the scene or mixture. Second, we select a target speaker among all speakers, and the model separates target speech with the target speaker's visual feature.

Decades ago, the idea of audio-visual speech separation had been explored by many non-deep methods [8]-[11]. Their main limitations are the inability to learn from large datasets and generalization to different speakers. Deep methods for audio-visual speech separation and speech enhancement have been proposed in recent years. Their main idea is to use deep visual features to assist the speech separation or enhancement. The Looking-to-Listen model [12] uses a pre-trained face recognition model to extract face embeddings from the face thumbnails as deep visual features and use large datasets further to learn the correlation between visual features and acoustic signals. Besides, the method in [13] uses face landmarks extracted with Dlib and achieve satisfactory performance. Unlike the methods mentioned above, many works [14]-[18] use lip region as the model's visual input directly. Lip movements are the most relevant but not the only information correlated to speech, and the facial movements obviously contain all speech-related information. In general, faces contain two kinds of information: identity-related and speech-related information. Compared with the former, the latter is a relative speaker-independent feature that is more suitable for speech separation tasks [19]. 

In this paper, inspired by the previous work of face generation [20], we propose a novel audio-visual speech separation model. Specifically, we utilize the adversarially disentangled method to obtain speech-related visual features from face thumbnails. Unlike the Looking-to-Listen model [12], the speech-related visual feature is extracted explicitly, which make our model achieve excellent results even on limited size datasets. Besides, compared with previous works [14]-[18], our model uses more complete speech-related visual information and may achieve better performance. We conduct experiments on 2-speaker mixture created from GRID [21] and TCD-TIMIT [22] audio-visual datasets, respectively, and use Signal-to-Distortion-Ratio improvement (SDRi) [23] as an evaluation measure. Results show that our proposed model outperforms the state-of-the-art audio-only model: Conv-TasNet [4] and three audio-visual models [12], [14]-[15]. We also provide two benchmark datasets (2-speaker mixture) created by us, which could be used to measure the effectiveness of audio-visual speech separation models.
%
\begin{figure}
\centerline{\includegraphics[width=0.8\columnwidth]{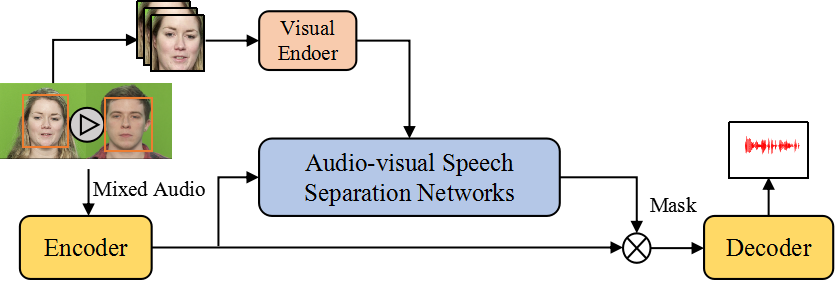}}
\caption{The framework of our proposed audio-visual speech separation model.}
\end{figure}
\section{The Proposed Model}
The framework of our proposed model is shown in Fig.1. First, we use the face detector to determine the number of speakers in the scene or mixture and obtain face thumbnails for each speaker. The attended speaker's speech-related visual feature is extracted by the visual encoder from corresponding face thumbnails, and we get mixed acoustic representation from mixed audio by speech encoder. The audio-visual speech separation networks take both mixed acoustic representation and target speaker's visual features as inputs for target mask prediction. After that, we do element-wise multiplication between mixed acoustic representation and mask to obtain the target speaker's acoustic representation, which could be decoded as target audio by the speech decoder. The same process is repeated for the remaining speakers to separate the sounds of all the speakers. We train visual encoder with an adversarially disentangle method. After completing the training, the visual encoder is kept frozen. The training method will be described in detail in Section \uppercase\expandafter{\romannumeral2}-A, and the implementation of audio-visual speech separation networks will be introduced in \uppercase\expandafter{\romannumeral2}-B.
\subsection{Adversarially Disentangled Method}
The method of adversarially disentangle could be viewed as two stages. First, as shown in Fig.2(a), we learn joint audio-visual representation from clean audio and video pairs and use three supervisions training loss to force the two embedded features ($F_v$ and $F_a$) to share the same distribution (\uppercase\expandafter{\romannumeral2}-A-1, \uppercase\expandafter{\romannumeral2}-A-2, and \uppercase\expandafter{\romannumeral2}-A-3).
Second, after the first stage is completed, we disentangle the speech-related visual feature from joint audio-visual representation with an adversarial training method (\uppercase\expandafter{\romannumeral2}-A-4).
The whole training procedure is summarized in \uppercase\expandafter{\romannumeral2}-A-5 finally.
\subsubsection{Sharing the Same Classifier}
After getting $F_v = [f_v^{(1)},...,f_v^{(n)}]$ and $F_a = [f_a^{(1)},...,f_a^{(n)}]$ from $E_v$ and $E_a$ respectively. We share the same classifier $C$ for doing visual and audio speech recognition task (word-level classification), which could enforce them share the same distribution. The supervision loss function is denoted as $\mathcal{L}_w$:
\begin{equation}
  \mathcal{L}_w = -\sum_{k=1}^{N_w}p_k(log(\hat{p}_k^v)+log(\hat{p}_k^a)),
\end{equation}
where $\hat{p}_k^v=softmax(C(F_v))_k$, $\hat{p}_k^a=softmax(C(F_a))_k$, $p_k$ is the true class label and $N_w$ is the total number of word label in train dataset.
\begin{figure}
\centerline{\includegraphics[width=\columnwidth]{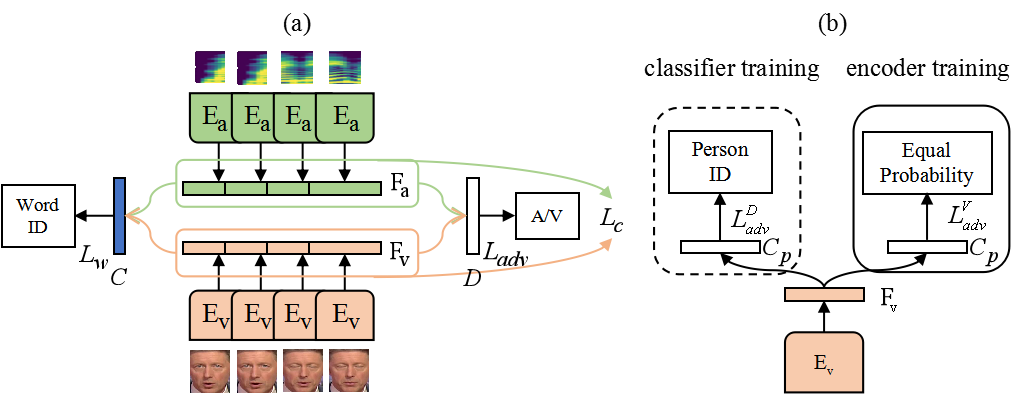}}
\caption{Illustration of adversarially disentangled method. (a): the method of learning joint audio-visual representation. The inputs of visual encoder $E_v$ and audio encoder $E_a$ are face thumbnails and mel-frequency cepstral coefficient respectively. (b): The procedure of adversarial training.}
\end{figure}
\subsubsection{Adversarial Training}

Adversarial training is an effective method to push the two distribution further closer. So we use a simple version adversarial training to further push the visual features $F_v$ and audio features $F_a$ to be in the same distribution. The discriminator $ D $ is a two-class (audio or video) classifier. The training process is described below. First, the visual encoder $E_v$ and audio encoder $E_a$ are frozen, and we train $D$ for distinguishing the source of the feature, the loss function is denoted as $\mathcal{L}_{adv}^1$. Then, the $D$ is frozen, and we train $E_a$ and $E_v$ to prevent the classifier from success, this loss function is denoted as $\mathcal{L}_{adv}^2$. $\mathcal{L}_{adv}^1$ and $\mathcal{L}_{adv}^2$ are defined as:
\begin{equation}
  \mathcal{L}_{adv}^1 = \|p_v-\sigma(D(F_v))\|_2^2+\|p_a-\sigma(D(F_a))\|_2^2,
\end{equation}
\begin{equation}
  \mathcal{L}_{adv}^2 = \|p_a-\sigma(D(F_v))\|_2^2+\|p_v-\sigma(D(F_a))\|_2^2,
\end{equation}
where $p_v$ equals to 0 represents the source is video, $p_a$ equals to 1 represents the source is audio, and $\sigma$ represents the $sigmoid$ function.

\subsubsection{Contrastive Loss}
We adopt the contrastive loss which aims to bring closer two embedded features. During training, the $m$-th and $n$-th samples are drawn with label $l_{m=n}=1$ and $l_{m\ne{n}}=0$ from a batch of $N$ audio-video pairs. The distance between $F_{v(m)}$ and $F_{a(n)}$ here is the Euclidean norm $d_{mn}=\|F_{v(m)}-F_{a(n)}\|_2$, and the contrastive loss function is denoted as $\mathcal{L}_c$:
\begin{equation}
  \mathcal{L}_c = \sum_{m,n=1}^{N,N}(l_{mn}d_{mn}+(1-l_{mn})max(1-d_{mn},0)).
\end{equation}
\subsubsection{Obtain Speech-related Visual Feature}
In this section, we describe how we disentangle the speech-related feature from the joint audio-visual representation with an adversarial training method. The training procedure is shown in the Fig.2(b). We freeze the visual encoder $E_v$ and train classifier $C_p$ for mapping $F_v$ to the $N_p$ Person-ID classes firstly. The loss function for training the classifier is softmax cross-entrogy loss, which is denoted as $\mathcal{L}_{adv}^D$:
\begin{equation}
\mathcal{L}_{adv}^{D} = -\sum_{j=1}^{N_p}p^jlog(softmax(C_p(F_v))_j),
\end{equation}
where $N_p$ is the number of personal identities, and $p_j$ is the true one-hot label. Then we update the visual encoder $E_v$ while freezing the classifier $C_p$. The way to ensure that the features have lost all identity-related information is that it produces the same prediction probabilities ($1/N_p$) for all classes after $F_v$ being sent into $C_p$. In other words, $E_v$ is trained to prevent the $C_p$ from success. The loss function for training the visual encoder is denoted as $L_{adv}^V$:
\begin{equation}
\mathcal{L}_{adv}^{V} = \sum_{j=1}^{N_p}\|{softmax(C_p(F_v))_j-\frac{1}{N_p}}\|_2^2.
\end{equation}
\begin{figure}
\centerline{\includegraphics[width=\columnwidth]{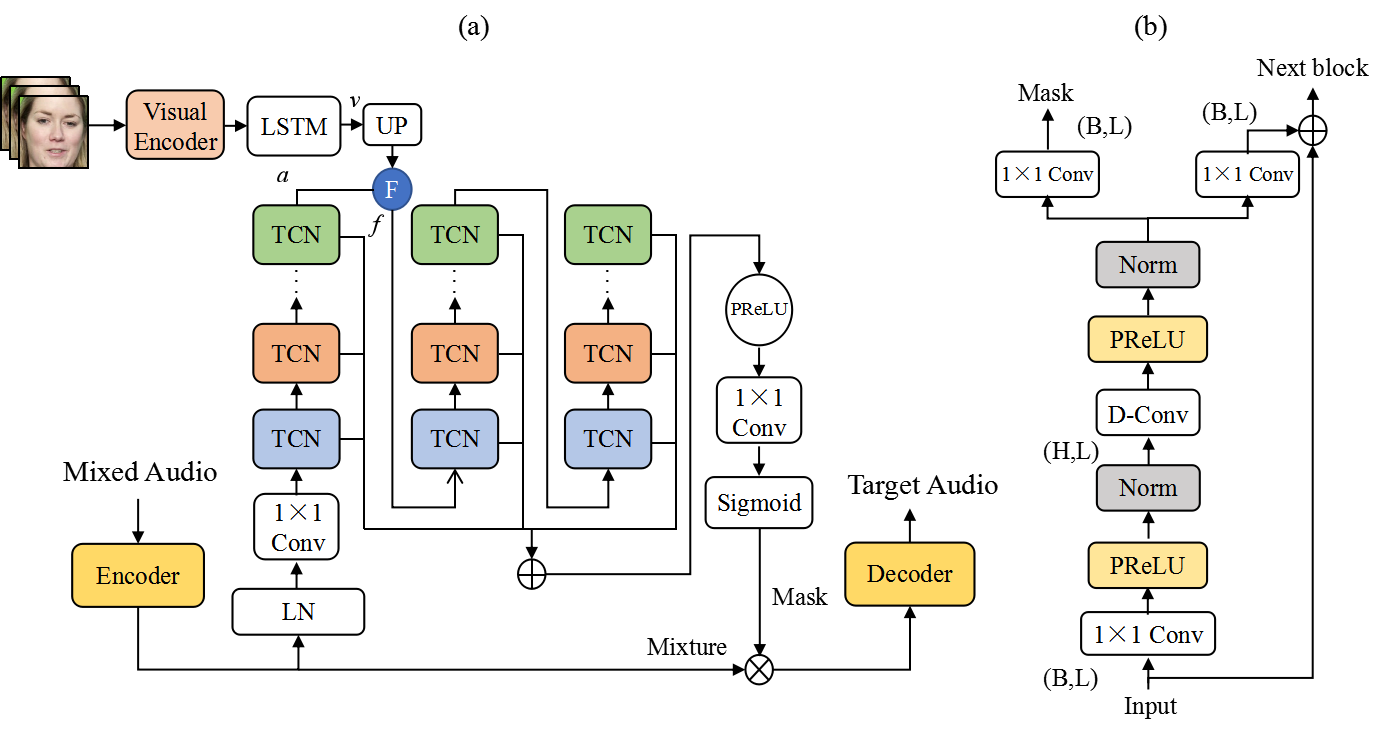}}
\caption{(a): The flowchart of audio-visual speech separation networks. \textbf{UP}: upsample, \textbf{F}: multimoda fusion, \textbf{LN}: layer normalization. (b): Details of TCN block.}
\end{figure}
\subsubsection{Details of Training Process}
The training process adopts the paradigm of adversarial training, which is as follows. For first stage, we freeze $E_v$, $E_a$ and $C$ and train $D$ with loss function $\mathcal{L}_{adv}^1$. Then, we freeze $D$ and train $E_v$, $E_a$ and $C$ with loss function $\mathcal{J}_1 = \mathcal{L}_{adv}^2+\mathcal{L}_c+\mathcal{L}_w$. For second stage, we freeze $E_v$, $E_a$ and $C$ and train $D$, $C_p$ with loss function $\mathcal{J}_2 = \mathcal{L}_{adv}^1+\mathcal{L}_{adv}^D$. Then we freeze $D$, $C_p$ and train $E_v$, $E_a$ and $C$ with loss function $\mathcal{J}_3 = \mathcal{L}_{adv}^2+\mathcal{L}_c+\mathcal{L}_w+\mathcal{L}_{adv}^V$.

\subsection{Audio-visual Speech Separation Networks}
Similar to the separation module in Conv-TasNet [4]. In our audio-visual speech separation networks, we stack the Temporal Convolutional Network (TCN) blocks by exponentially increasing the dilation factor to capture the long-range dependency of the speech signal, as shown in Fig.3(a). The TCN block has been proven effective in speech separation tasks [4] and details of it is shown in Fig.3(b). In our model, we use three groups of TCN blocks and eight TCN blocks in each group.

For visual pathway, the speech-related visual features $v_1\in{\mathbb{R}^{T_1\times{N{_1}}}}$ is extracted by visual encoder firstly, then we use 3-layers BiLSTM to capture the temporal dependency in it and obtain deep visual features $v\in{\mathbb{R}^{T_1\times{N_{2}}}}$. For audio pathway, the mixed audio $a_1\in{\mathbb{R}^{1\times{T}}}$ is encoded to mixed acoustic representation $a_2\in{\mathbb{R}^{D\times{T_2}}}$ by encoder (convolutional layer). The mixed acoustic representation $a_2$ is normalized firstly, then pass through a $1\times1 Conv$ with $B$ filters and obtain $a_3\in{\mathbb{R}^{B\times{T_2}}}$. The first group of TCN blocks take in $a_3$ and output deep audio features $a\in{\mathbb{R}^{B\times{T_2}}}$. At the same time, we fuse deep visual features $v$ and deep audio features $a$ by fusion module, which will detailed in \uppercase\expandafter{\romannumeral2}-B-1. In this way, we have the fusion features $f\in{\mathbb{R}^{B\times{T_2}}}$, which serves as the input of the next two groups of TCN blocks. Finally, we do element-wise multiplication between mixed acoustic representation and mask $m\in{\mathbb{R}^{D\times{T_2}}}$ generated by model and obtain masked (target) acoustic representation which is decoded as target audio by decoder (transposed convolutional layer). The second $1\times1 Conv$ with $D$ filters ensure that the same number of channels between mask and mixed acoustic representation, and sigmoid function contrains the element in mask between 0 and 1. All normalizations in the networks are global layer normalization (gLN) [4].

\subsubsection{Multi-modal Fusion}
First, to synchronize the time resolution between deep audio features $ a $ and deep visual features $v$, the upsample operation is done on the latter. Then they are concatenated over the channel dimension. A linear layer $\mathcal{P}$ is adopted to reduce the channel dimensions of the concatenated features and get fusion features $f$. The description above could be denoted as: 
\begin{equation}
f=\mathcal{P}([a;upsample(v)]).
\end{equation}
\subsubsection{Loss Function}
The objective of training the separation networks is minimizing the negative scale-invariant source-to-noise ratio (SI-SNR), and SI-SNR is defined as follows:
\begin{equation}
  \mbox{SI-SNR} = 10log_{10}\frac{\|{\alpha{\cdot{a_t}}}\|^2}{\|{a_e-\alpha{\cdot{a_t}}}\|^2},
\end{equation}
where $a_e$ and $a_t$ are estimated audio and target audio respectively, and they are normalized to zero mean. Besides, $\alpha=a_e^Ta_t/\|{a_t\|^2}$.
\section{Experiments}
\subsection{Datasets}
The visual encoder is trained on the LRW dataset [24] and the MS-Celeb-1M dataset [25]. We use the 2-speaker mixture of the GRID and TCD-TIMIT audio-visual datasets to test the performance of speech separation of our proposed model, and the way of constructing datasets is the same as [14]. 

Regarding the GRID corpus, the dataset contains 18 male speakers and 15 female speakers, and each of them has 1000 frontal face video recordings (“s21” has to be discarded). The length of each video is three seconds. We randomly select 3 males and 3 females to construct a valid set of 2.5 hours and another 3 males and 3 females for a test set of 2.5 hours. The rest of the speakers form the training set of 30 hours. To construct a 2-speaker mixture, we randomly choose two different speakers first, randomly select audio from each chosen speaker, and finally mix two audios at a random SNR between -5 dB and 5 dB. The corresponding two videos are concatenated to simulate a cocktail party scene, as shown in Fig.1.

The TCD-TIMIT corpus consists of 59 speakers (32 males and 27 females), and each speaker reads 98 sentences from the TIMIT [26] corpus, resulting in durations of around 5 seconds for each video. We randomly select 6 males and 5 females to construct a validation set of 2.5 hours and another 6 males and 5 females for a test set of 2.5 hours. The rest of the speakers form the training set of 30 hours. The data generation process is similar to that of the GRID dataset.
\subsection{Baselines and Our T-F Domain Model}
The audio-visual baselines to be compared include Looking-to-Lisen model\footnote{https://github.com/JusperLee/Looking-to-Listen-at-the-Cocktail-Party} [12], AVDC [14] and AV-Match [15]. Besides, we also compare our method with state-of-the-art audio-only baseline Conv-TasNet\footnote{https://github.com/naplab/Conv-TasNet} [4]. The Looking-to-Listen and Conv-TasNet model obey the best configurations in their paper. Because three audio-visual baselines are all time-frequency (T-F) domain models, we also implement our model in the T-F domain.
In our T-F domain model, we use the Short-Time Fourier Transform (STFT), and Inverse Short-Time Fourier Transform (ISTFT) replace encoder and decoder. The mixed spectrogram $\mathcal{X}\in{\mathbb{R}^{F\times{T_3}\times2}}$ is obtained by do STFT on mixed audio $a_1$ firstly. Then we concatenate the real and imaginary part of $\mathcal{X}$ over the frequency channel and get $\mathcal{X}_1\in{\mathbb{R}^{2F\times{T_3}}}$, which as the input of our model. The output is complex Ratio Mask (cRM), which is represented by $M\in{\mathbb{R}^{F\times{T_3}\times2}}$ . When masking with cRM, target audio is obtained by performing ISTFT on the complex multiplication of the predicted cRM and mixed spectrogram. We define the loss function as the mean squared error (MSE) between the clean spectrogram and the estimated spectrogram.

\subsection{Setup}
\subsubsection{Video and Audio Pocessing}
For each video clip, we resample the video to 25 FPS and convert it to video frames firstly. Then we use a face detector (MTCNN [27]) to find faces in each frame and resize the face image to $256\times256$, as shown in Fig.1. Besides, the audio is resampled to 16 kHz. For our T-F model, STFT is computed using a Hann window of length 25 ms, hop length of 10 ms, and FFT size of 512. We follow the implementation in [28] to extract the $mfcc$ features.
\subsubsection{Model and Training Details}
The network architecture of visual encoder $E_v$ and audio encoder $E_a$ are similar to [20]. The parameters for audio-visual speech separation networks $B$, $H$, $N_1$, $N_2$, and $D$ are set to be 128, 512, 256, 128, and 512. Besides, the kernel size and stride of encoder and decoder are both set to 16 and 8. We implement the whole model using Pytorch, the batch size for pre-training visual encoder is set to be 18 with 1e-4 learning rate, and using Adam algorithm [29]. The audio-visual speech separation networks are also optimized by Adam algorithm. The learning rate begins with 1e-3 and halves when the loss increases on the validation set for at least 3 epochs. An early stopping scheme is applied when the loss increased on the validation set for 10 epochs. The batch size is set to 8, and gradient clipping with a maximum $L_2$-norm of 5 is applied during training.

\subsection{Experimental Results}
The experimental results of ours and other models are summarized in Table \uppercase\expandafter{\romannumeral1}. We could find several interesting conclusions from it.

First, our T-F domain model performs significantly better than three audio-visual models on GRID and TCD-TIMIT audio-visual datasets, such as Looking-to-Listen (L2L) [12], AV-Match [15], and AVDC [14]. It should be noted that in order to make a fair comparison with the Looking-to-Listen model, we run their model on our created benchmark dataset. The main reason that the Looking-to-Listen model performs worse than ours: as we hypothesized, their method is difficult to learn the speech-related information in a dataset of limited sizes, such as GRID. Besides, compared with the AVDC, our model (T-F) achieves an improvement of 1.83 dB and 3.86 dB on the GRID and TCD-TIMIT datasets, respectively.

Second, compared with our baseline Conv-TasNet, our time-domain model surpass its performance on GRID and TCD-TIMIT dataset, respectively. Besides, there is another advantage that our model is more flexible in practical applications: we could obtain the voice of any person in the scene that contains any number of speakers. Compared with the audio-only model DC, the performance improvement of the AVDC model is 1.16 dB and 1.15 dB on GRID and TCD-TIMIT datasets, which could be viewed as a result of introducing visual information. It is important to note that the AVDC model uses a two-stage multimodal fusion strategy and additional optical flow information. By contrast, our model uses a simple multimodal fusion strategy (concatenate) gains 1.34 dB and 2.01 dB on these two datasets, respectively, which reflects the effectiveness of our proposed visual feature extraction method.
\begin{table}
  \caption{SDR improvements with different model based on 2-speaker mixtures created from GRID and TCD-TIMIT datasets. '*' indicates our training, using published code and our created dataset.}
  \label{tab:example}
  \centering
  \renewcommand{\arraystretch}{1.2}
  \begin{tabular}{ccc}
  \hline
  \hline
  Datasets & Method & SDRi (dB)\\
  \hline
  \multirow{7}{*}{GRID}  & DC [14] & 7.72 \\
           & AV-Match [15] & 8.11 \\
           & L2L* [12] & 8.31 \\
           & AVDC [14] & 8.88 \\
           & \textbf{Ours} (T-F) & 10.71 \\
  \cline{2-3}
           & Conv-TasNet* [4] & 14.40 \\
           & \textbf{Ours} & \textbf{15.74} \\
  \hline
   \multirow{5}{*}{TCD-TIMIT}        & DC [14] & 6.32 \\
           & AVDC [14] & 7.47 \\
           & \textbf{Ours} (T-F) & 11.33\\
  \cline{2-3}
           & Conv-TasNet* [4] & 14.17 \\
           & \textbf{Ours} & \textbf{16.18} \\
  \hline
  \hline
  \end{tabular}
\end{table}
\section{Conclusions}
In this paper, we propose a novel audio-visual speech separation model that adopts an adversarially disentangled method to extract speech-related visual features from visual inputs and use it to assist the speech separation. Results show that our model achieves excellent performance in a cocktail party setting. In addition, our model is suitable for scenes where human's visual cues could be obtained with high quality, such as video conferencing, video calling, and multi-modal robot interaction.

.
\end{document}